# "HEALTHWISE" – AN ANDROID APPLICATION FOR PERSONAL HEALTH AND NUTRITION MANAGEMENT

Hamsa Shwetha V[1], Kruthika Prarthana M K[2], Shivani C[3], Siva Naga Suresh Purama[4], S P Muthukumar[5]

[1,2,3] Department of Computer Science and Engineering, [4,5] Computer Centre, Biochemistry & Nutrition
[1,2,3] V.V.C.E, Mysore, India, [4,5] Central Food Technological Research Institute, Mysore, India
hamsashwetha@gmail.com

*Abstract*— Now-a-days, people are getting more health conscious and tend to keep a check on nutritional gain from the packed food products they consume. This application- christened "healthWISE" – helps a mobile user to scan/enter the barcode on the packed food product, know the nutritional information, and upon entering the quantity to be consumed, it displays the energy the individual can consume. If the user has exceeded the stipulated amount of calories, then it suggests the exercise to burn the extra number of calories.

**Keywords-Android; nutrition**

## I. INTRODUCTION

Technology is evolving at a faster pace, than it was a decade ago. With the advent of hand-held mobile devices, technology has reached all segments of population. The device manufacturers are striving hard to enrich the user experience with easy-to-use mobile phones, tablets etc. There are a number of applications that are being developed to run on these devices. To back these applications, device needs a powerful Operating System. One such OS which is the major technological breakthrough is Android.

The increase in number of mobile phones having Android OS has resulted in increase in number of third-party applications. There are number of applications which have brought together Technology and many other fields. One such application tried for effective coalesce of Technology and Nutrition is "healthWISE".

In India, the numbers of mobile phone users are increasing at an enormous rate. As Android became popular, there is a radical shift in the mobile phone market. On the other hand, users have become more health-conscious and dieticians or nutrition experts are gaining prominence. Nevertheless, people care about their family's health.  So to converge all these paths into one, it seemed that if a user can get information about a food product that the user comes across in a supermarket, a suggestion that can help to make decision whether to buy the product and use it or not. This can be done using a mobile phone supported with Android. This paper describes the process and result of such an application "healthWISE" – which has been named so to signify main intentions (a) Application helps to protect user's health wisely   (b) Application in terms of health.

## II. LITERATURE SURVEY

### A.  Android Architecture

Android platform has a software stack with operating system, middleware and key applications. The layers are the kernel, application framework layer, applications layer and the libraries. The application layer is the location where all applications are present. The application framework layer provides APIs and managers to help the developer exploit the functions of Android [1]. Android applications are written in the Java programming language. The Android SDK tools compile the code—along with any data and resource files—into an Android package, an archive file with an .apk suffix. All the code in a single .apk file is considered to be one application and is the file that Android-powered devices use to install the application.

Once installed on a device, each Android application lives in its own security sandbox: The Android operating system is a multi-user Linux system in which each application is a different user [2].

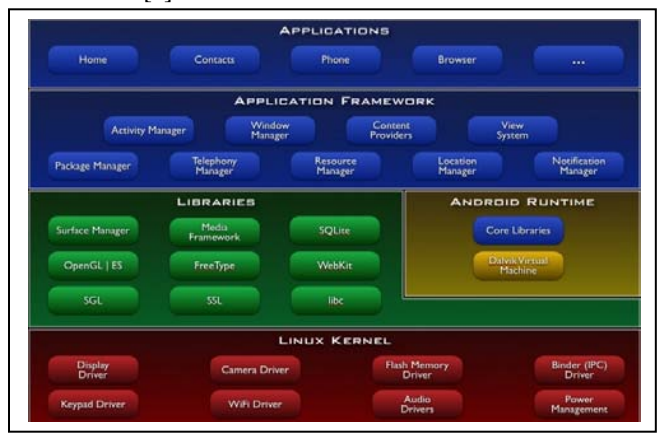

Fig 1 Android Architecture

### B.  Why Android?

The ease of development and deployment on to a mobile phone is the main reason of choosing Android to deploy the application.

III. DESIGN

This application works as described in Fig 2: When the user installs "healthWISE", user needs to create own profile and family's profile. After profile creation, user can update or delete any of the profiles.

In the supermarket, user comes across a packed food product. Application can be used to capture barcode. If the lens is unable to focus barcode due to bad light or any other reasons, user can enter the barcode manually in the space provided. Mobile fetches nutritional information of the product from the server. Since this data is huge, it is stored in the server called as Nutrition Server. If the user wishes to buy it, user can check if it suits the calorie requirements for that day. Again the mobile interacts with server to get suggestions and displays it on the screen.

This application has been divided into three modules: Requirement module, Nutrition module, and Exercise module.

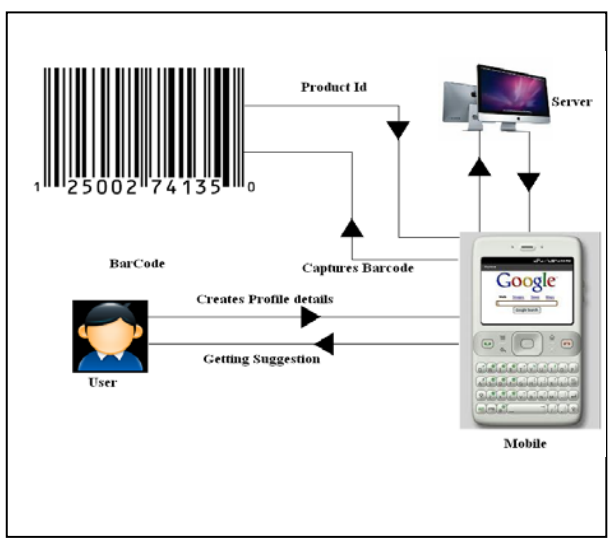

Fig 2.1 healthWISE design

*A. Requirement Module*

This module contains functions to create user profiles, update user profiles and store them in phone's storage area. When the user installs the application, welcome screen is displayed with options to create a profile [Table 1]. User can do so by using a form provided and is stored in phone's Database. Once the profile is created, there is an option to update or delete the profile. For every action, suitable toast messages are popped up to help the user. The database contains tables such as Nutrition Requirements for an individual [5-10], product information, a daily log to keep log of user's consumption information and exercise chart – that contains suggested exercises to burn calories.

*B. Nutrition Module*

This module contains functions to capture/enter barcode, retrieve information, check the amount energy the user can consume in a day, and add it to the database.

After the profile is created, user can capture the barcode. After the barcode is captured, if the product is stored in database, then its details are fetched from the server using Web Services. The protocol used is Simple Object Access Protocol (SOAP). SOAP makes object access simple by allowing applications to invoke object methods or functions, residing on remote servers. A SOAP application creates a request block in XML, supplying the data needed by the remote method as well as the location of the remote object itself. Though Android SDK does not provide SOAP library, but ksoap2 library available online [3], which is easy to understand and implement, can be included. The suggestions are displayed in green colour, to signify that user is within allowed energy limits and user can consume the product and if the suggestions are in red colour, it signifies user has exceeded the limit [Table 2]. So the user can decide whether to buy the product or not, for own consumption or any of family members. Time of consumption is also provided – breakfast, lunch or dinner. In future, expecting that most of the food products come with bar-coded packets in India, this can work very effectively. When user clicks 'add' button indicating that food was consumed, then it is added into the daily log maintained in the server. Mail is sent to the user, provided at the time of registration, about the product consumed and the time of consumption.

*C. Exercise Module*

This module has functions to store the exercises and the calories they burn, and fetch proper exercise to burn the calories which the user has gained in excess.

IV. IMPLEMENTATION

Since the application is to be developed for Android, we make necessary environment set-up and implement the application as per design.

*A. Environment Set-up*

*a)* To develop an application, an Integrated Development Environment (IDE) is required. The choice selected is Eclipse Indigo IDE.

*b)* For an Android application development, Android SDK and Android Virtual Device(Emulator) are essential. [4]

*c)* A new android project is created to develop layouts and code the business logic for the same.

*d)* A Java project is created to code the web services part and for accessing the database.

e) To develop this application, we mainly import an open source library "ZXing". ZXing (pronounced "zebra crossing") is an open-source, multi-format 1D/2D barcode image processing library implemented in Java, which ports to other languages. This library currently supports these formats: UPC-A and UPC-E, EAN-8 and EAN-13, Code 39,

Code 93, Code 128, QR Code, ITF, Coda bar, RSS-14 (all variants), Data Matrix, PDF 417 ('alpha' quality), Aztec ('alpha' quality).

*B.  Implementation*

Once the necessary environment is set-up, coding is done as per design. Android layouts can be done either graphically or using XML. The required business logic for the respective layout is coded using Java Class. To use any of the phone's features such as Camera, Internet/Wi-Fi etc., permissions are obtained by mentioning in AndroidManifest.xml file.

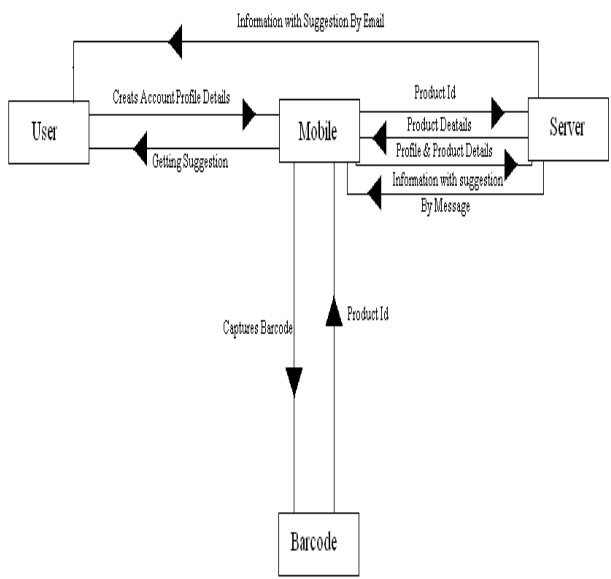

Fig 2.2  Data Flow Diagram

The web services part is coded by importing ksoap2 library and using the APIs provided.

*C.  Testing*

The application is tested by launching the Android Virtual Device of required version of Android. This emulates the device and runs the application just as it would on a real phone. The emulator can be seen in Fig 3.

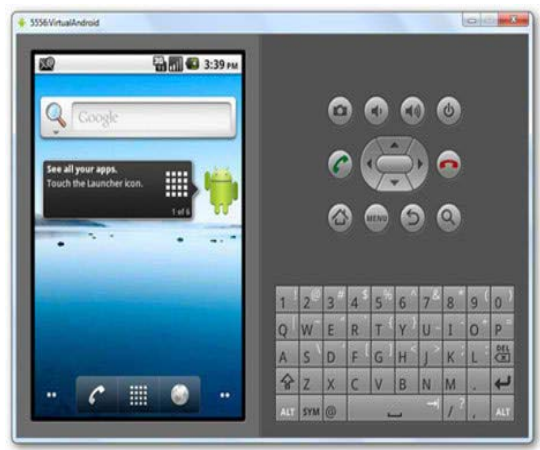

Fig 3.1   Android Emulator

Fig 3.2 shows the Welcome Screen, which is also called the Splash screen. It is the screen which is popped up when application launches and never shows up if back button is pressed. That is, it displays only once during the running of application. It displays for few seconds and moves to next screen or it can be changed by touching the screen.

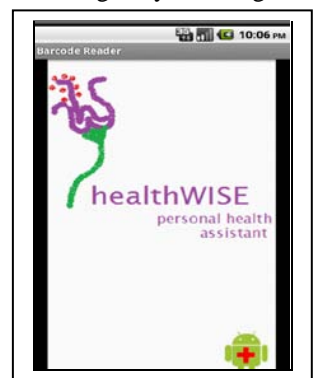

Fig 3.2  Welcome Screen

Main Menu is as shown in Fig 3.3. The Capture Barcode and Update profile buttons are disabled until the user creates profile. Create Profile and Update profile screens navigate to a screen in Fig 3.3(right), which contains a form to fill up details such as age, height, weight, activity and email. Activity helps to decide the calories required for an individual. Email is taken to send mail to user regarding the consumption details. This is done so as to have complete record of how much a user has consumed.

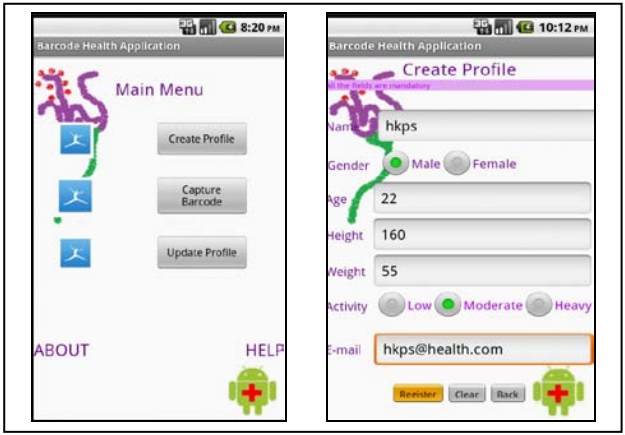

Fig 3.3 Main menu Screen and Profile Creation Screen

Next screen shown in Fig 3.4 contains options to go to Camera to capture barcode. Or the user can choose to enter the barcode. Fig 3.4 (right) displays user names for whom user wishes to buy the product. Along with that it also displays product information.

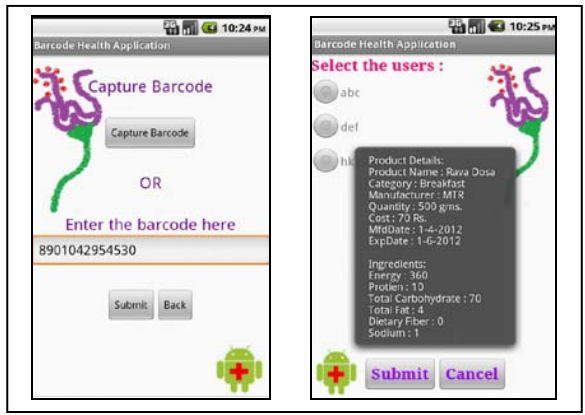

Fig 3.4 Capture/Enter Barcode Screen and Product Details screen

Fig 3.5 displays user details and quantity of consumption. Energy break-up for breakfast, lunch and dinner are displayed. And when user presses check button, green light is displayed to show user is within the energy limits as in Fig 3.6 and red light to show he has exceeded the daily energy requirements, and a set of exercises to burn the extra calories is displayed as in Fig 3.6.

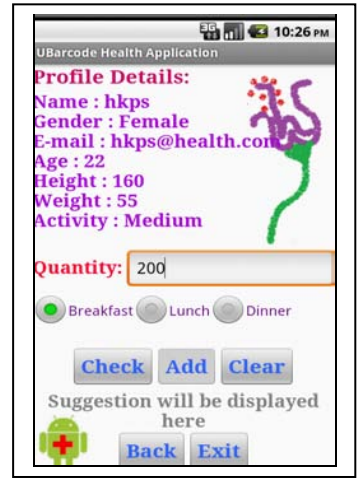

Fig 3.5 Profile Details Screen

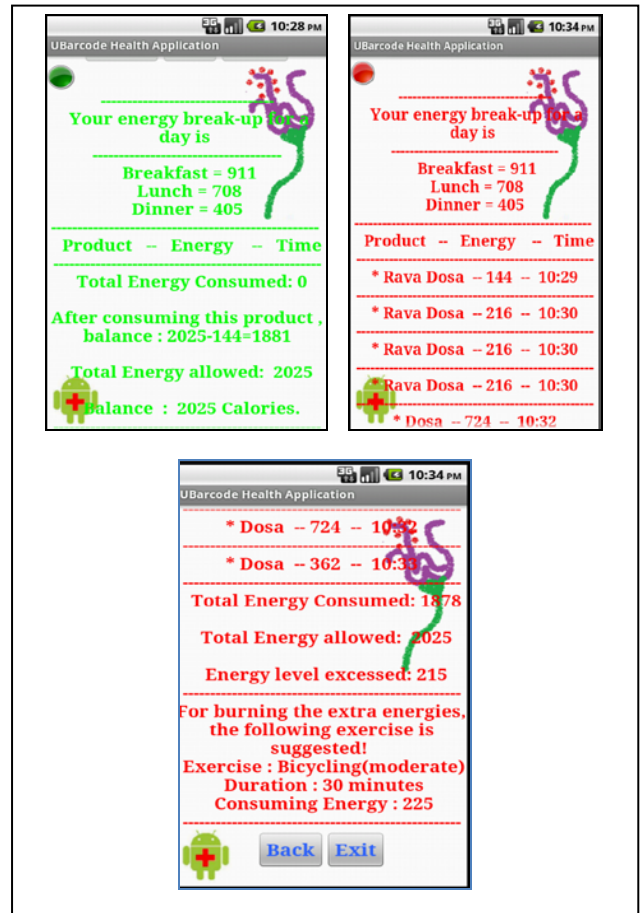

Fig 3.6 Energy Check screen and Energy exceed screen

## V. RESULTS

The application is deployed onto the mobile phone by using "export application to Android phone" which provides an .apk file (application package file). This is uploaded to phone and is installed. Then the application is run. The results for the following inputs have been tabulated:

Table 1   User Input

| SI No | User Input | | | | | | |
|---|---|---|---|---|---|---|---|
| | Gender | Age | Height (cms) | Weight (kgs) | Standard energy (kCal) | Activity | Required Energy (kCal) |
| 1 | Male | 20 | 170 | 60 | 2200 | High | 2750 |

Table 2   Output in different trials

| Trial | Output | | |
|---|---|---|---|
| | Energy Consumed (kCal) | Balance (kCal) | Results |
| 1 | 1500 | 1250 | Suggestions displayed in green color text. The user is within Energy limits |
| 2 | 1000 | 250 | Suggestions displayed in green color text. The user is within energy limits |
| 3 | 500 | -250 | The user has exceeded the limits by 250. The suggestions are displayed in red color text. Exercises to burn extra calories are displayed |

ACKNOWLEDGMENT

The authors wish to thank Director, CFTRI. We also thank with deep gratitude, Sri. S. N. Krishna Rao, Head, Computer Centre, CFTRI, for giving us an opportunity to work on this research work at CFTRI. We also wish to thank Our Head of Department of CS, Dr. Mahesh Rao for encouraging us in carrying out this project and our internal guide, Mrs. Sunita B K, Assistant Professor, for her support in every stage of this research work.